\documentclass[prb,twocolumn,nofootinbib,floatfix]{revtex4-1}
\usepackage{xcolor}
\usepackage{graphicx}

\usepackage{ifpdf}
\ifpdf
   \usepackage[unicode=true,colorlinks=true]{hyperref}
\fi

\begin{document}

\title{Tight-binding calculations of SiGe alloy nanocrystals in SiO$_2$ matrix}

\author{A.\,V. Belolipetsky, M.\,O. Nestoklon, and I.\,N. Yassievich }
\address{Ioffe Institute, 194021 St. Petersburg, Russia}


\begin{abstract}
In the empirical tight-binding approach we study the electronic states in 
spherical SiGe nanocrystals embedded in SiO\textsubscript{2} matrix.
The energy and valley structure is obtained as a function of Ge composition 
and nanocrystal size. 
The calculations show that the mixing of hot electrons in the nanocrystal 
with the electrons in wide band gap matrix is possible and this mixing 
strongly depends on the Ge composition in the nanocrystal.
\end{abstract}

\maketitle

\section{Introduction}

The SiGe alloy has the crystal structure similar to the bulk silicon and it is possible to 
create the Si$_{1-x}$Ge$_x$ layers on the silicon substrate changing the content of Ge in 
the wide range from $x = 0$ (pure silicon) to $x = 1$ (pure germanium).
This system is widely used in modern electronics: by changing the Ge content it is possible
to switch the electronic band structure from bulk silicon to germanium. An important feature 
of Si$_{1-x}$Ge$_x$ material is high mobility and convenient, well-developed production technology. 
These properties have opened a way of using Si$_{1-x}$Ge$_x$ to manufacture different electronic 
devices: high-speed transistors,\cite{Meyerson} long-wave infrared detectors,\cite{Lin}
solar cells based on p-i-n structures.\cite{Bidiville}

The use of silicon and germanium in optoelectronics is limited by the fundamental feature of their indirect 
band structure: the extrema of the conduction band are near the edge of the Brillouin zone, and the top of 
the valence-band is at the center of this zone. However, in nanocrystals (NCs), electrons and 
holes are localized and no longer have a definite (quasi)momentum due to the Heisenberg uncertainty 
principle. It stimulates the rapid development of technology, experimental, and theoretical work on the study 
of optical phenomena in silicon and germanium nanoparticles. There is a large number of reviews and 
monographs devoted to the successful application of silicon nanocrystals in SiO$_2$ matrix 
in optoelectronics, photovoltaics and medicine (see Ref.~\onlinecite{Priolo}).

In Ref.~\onlinecite{Takeoka} the photoluminescence from spherical Si$_{1-x}$Ge$_{x}$ alloy nanocrystals 
with 4--5~nm diameter in SiO$_{2}$ matrix was studied as a function of the Ge content. 
The samples were fabricated by the cosputtering of Si, Ge, and SiO$_{2}$ 
and postannealing at 1100C. 
Recently, similar NCs prepared using the same technique were studied in details by spectroscopy and 
time-resolved spectroscopy.\cite{Ngo1,Ngo2}

In the present paper, we develop the empirical tight binding (ETB) approach\cite{Jancu98} for the 
modeling of electronic states 
and energy levels of SiGe nanocrystals embedded in SiO$_2$ matrix and present the results of modeling for 
Si$_{1-x}$Ge$_x$ nanocrystals with germanium content $x$ in wide region from $x=0$ to  $x=1$. 
The distributions of the density of states in reciprocal real and real space shows the X--L valley 
transition as a function of Ge content and also shows that excited electronic states with 
relatively small energy may strongly penetrate the SiO\textsubscript{2} matrix.



\section{Virtual crystal approach for SiGe alloy}
To simulate the alloy in empirical tight-binding (ETB), two approaches are widely used. 
First, one may consider the randomly chosen atom distribution within 
the structure and then average over realizations.\cite{Wei90}
Second option is to use the virtual crystal approximation (VCA) which is the ETB 
parametrization of the averaged band structure of the alloy. 
Second approach has its limitations: It is valid only in the cases when 
one wants to neglect the effects due to the disorder in the alloy material. 

Recently it has been shown\cite{Nestoklon16} that for the VCA in ETB 
it is not necessary to parametrize the alloy. At least for ternary 
alloys in the InGaAsSb and AlGaAs family the band structure of an alloy 
may be constructed from the ETB parameters of basic materials. 
Here we adopt the same procedure for the SiGe alloy.
The parameters of the alloy are found 
from the parameters of Si and Ge: 
The lattice constant of the alloy is found as a linear 
interpolation between binaries (Vegard’s law). 
Then the parameters of Si$_{1-x}$Ge$_{x}$ alloy 
are found as a linear interpolation of ETB parameters 
of Si and Ge \textsl{strained to the lattice constant of the alloy}. 

In the method proposed in Ref.~\onlinecite{Nestoklon16} it is assumed that in the 
parametrization of binary materials is critically important to have accurate 
description of the deformation potentials of basic materials, in which case 
the band structure of alloy will also be precise. 
The parametrization for bulk Si and Ge (as well as for SiGe bonds) which
takes into account the change of the parameters due to strain may be found in 
Ref.~\onlinecite{Niquet}. 

In the ETB we follow the procedure of Ref.~\onlinecite{Jancu98} and construct the 
matrix elements of the tight-binding Hamiltonian using the standard 
procedure.\cite{Slater54}
The parameters also depend on the strain: 
First, the transfer matrix elements of bulk Si and Ge are scaled due to the 
change of the lattice parameter following standard generalized Harrison law\cite{Jancu98}
\begin{equation}
  V_{m;ijk} = V_{m;ijk}^0 \left( \frac{a_{m}}{a_{m}^0} \right)^{n_{m;ijk}}
\end{equation}
where $V_{m; ijk}^0$ and $a_{m}^0$ the transfer parameters and lattice constant  
for the material $m$ (Si, Ge or SiGe), and $n_{m;ijk}$ is the 
power in generalized Harrison law, these parameters are taken from Ref.~\onlinecite{Niquet}.
The lattice constant of the alloy is found from the linear interpolation
\begin{equation}
a_{Si_{1-x}Ge_{x}} = (1-x)a_{Si} + xa_{Ge} \;.
\end{equation}

The valence band offset (VBO) of the alloy depends linearly on $x$
\begin{equation}
E_{VBO}\left(Si_{1-x}Ge_{x}\right)=E_{VBO}\left(Ge\right)\cdot x
\end{equation}
where $E_{VBO}\left(Ge\right)=0.68eV$ (see Ref.~\onlinecite{Niquet}).

In addition to the change of transfer matrix elements we also take into account the 
shift of orbital energies proportional to hydrostatic component of strain tensor 
\begin{equation}
  E_{m;\beta} = E_{m;\beta}^0 + \alpha_{m;\beta} 3  \left( \frac{a_{m}}{a_{m}^0}-1 \right)
\end{equation}
where parameters $\alpha_{\beta}$ ($\beta$ indexes the basis function of ETB)
are also taken from Ref.~\onlinecite{Niquet}.

The splitting of diagonal energies is proportional to the off-diagonal components
of the strain tensor introduced in Refs.~\onlinecite{Nestoklon16,Niquet} is 
irrelevant to the construction of VCA alloy parameters and we do not discuss
it here.

\begin{figure}
\includegraphics[width=\linewidth]{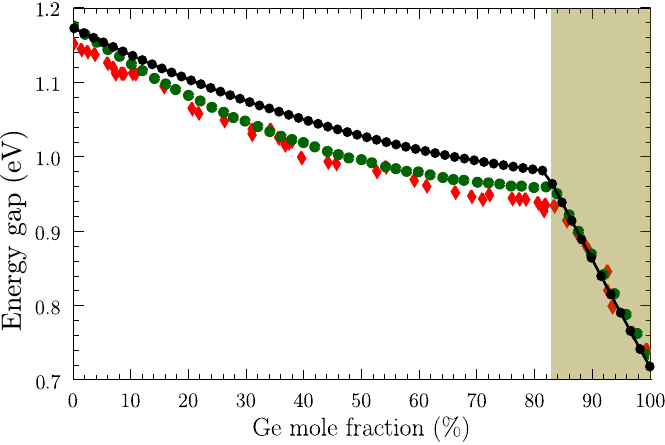}%
\caption{
  Dependence of the band gap energy on the Ge mole fraction. In black solid line 
  with dots we show the results of modeling in this study, in round (green online) 
  dots we show the results from Ref.~\onlinecite{Niquet}; in diamond shape (red online) markers we 
  show the experimental results.\cite{Dismukes} Shaded area shoes the 
  Ge contents where the minimum of the conduction band lies at points L.
}\label{fig:VCA}
\end{figure} 

We check that the constructed ETB parameters of SiGe alloy are in a good 
agreement with the results obtained using the random alloy description 
which may be found in Ref.~\onlinecite{Niquet}, see Fig.~\ref{fig:VCA}.

\section{Virtual crystal approach for SiO$_2$ matrix}

\begin{figure}
\includegraphics[width=\linewidth]{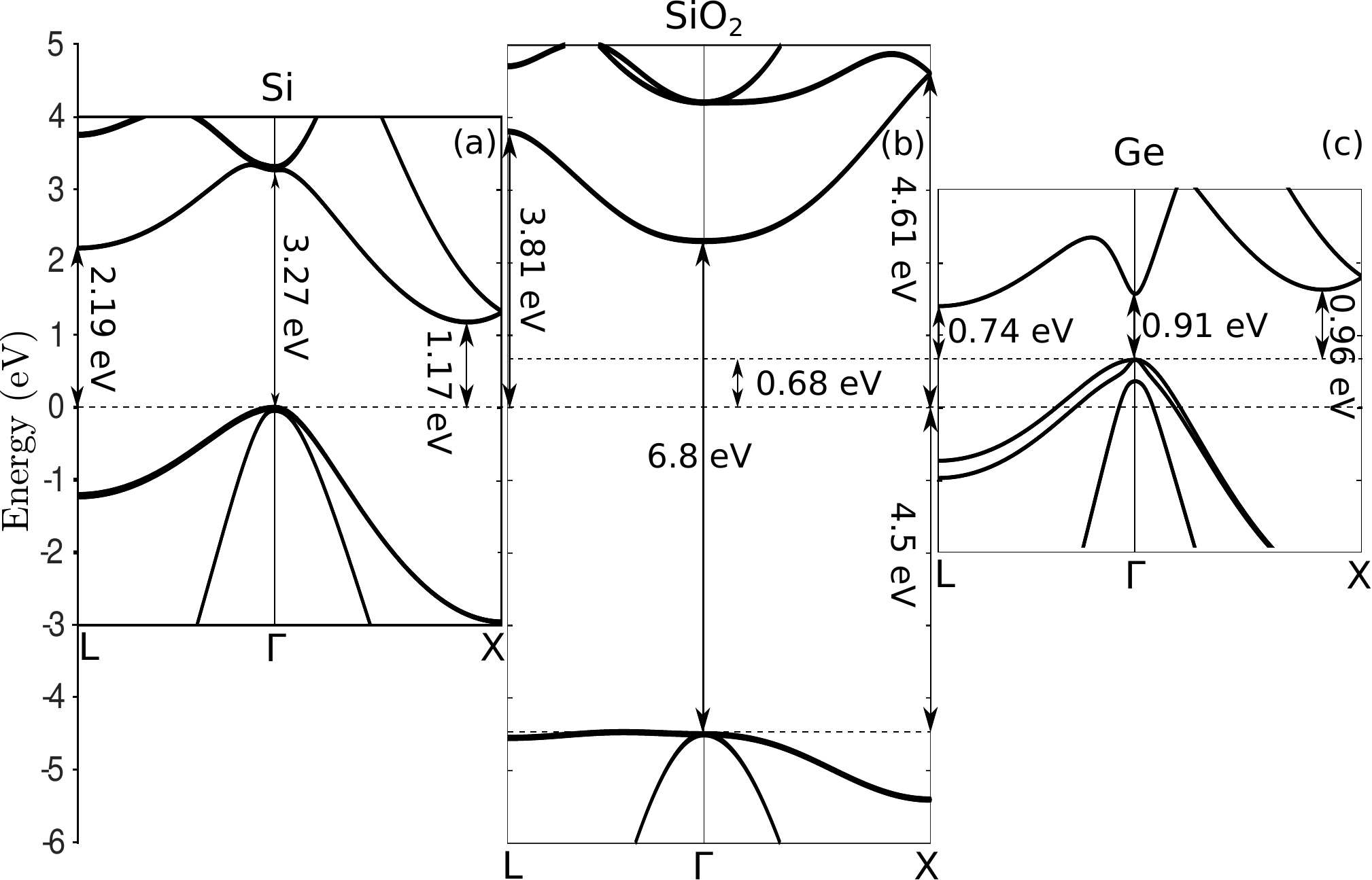}%
\caption{
  Band structures of silicon (a), $\beta$ -crystobalite (b) and  
  germanium (c) neer band gap.
}\label{fig:BS}
\end{figure} 
The simulation of Si$_{1-x}$Ge$_{x}$ alloy nanocrystals in SiO$_{2}$ by the 
tight-binding method is complicated by the fact that SiO$_{2}$ is an 
amorphous material. However, at the interface between SiO$_{2}$ and 
Si$_{1-x}$Ge$_{x}$ NCs, there is a large band offset for both types 
of carriers and, therefore, the electron and hole wave functions 
rapidly damp in the matrix. In this case, the role of disorder 
may be neglected and the most important factors are the general band 
structure of material surrounding nanocrystal and the boundary conditions 
between Si$_{1-x}$Ge$_{x}$ alloy and SiO$_{2}$.
It opens the possibility to simulate SiO$_{2}$ as a virtual 
crystal.\cite{Nestoklon16_Ge} We construct the virtual crystal 
with a band structure close to the band structure of $\beta$-cristobalite 
near the band gap edges, following Ref.~\onlinecite{Belolipetskiy}. 
The $\beta$-cristobalite  is the only polymorphous modification with a cubic lattice 
among SiO$_2$ crystals. As a target band structure we use the band structure
of $\beta$-cristobalite calculated from the first principles in Ref.~\onlinecite{Ching}.
As there is no strain of NCs in amorphous matrix,  
we set the lattice constant of the virtual crystal matching the lattice 
constant of the bulk Si$_{1-x}$Ge$_{x}$ alloy which forms the NC. 
The tight-binding  parameters for the virtual crystal SiO$_{2}$ may 
be found in Ref.~\onlinecite{Belolipetskiy} and in appendix.

Fig.~\ref{fig:BS} demonstrates the energy position of the band edges 
of the bulk silicon, virtual crystal SiO$_{2}$ and germanium.
The virtual crystal SiO$_{2}$ is the direct-band material with the 
band gap equal to 6.79~eV 
matching the value in $\beta$-cristobalite, the extrema of 
conduction and valence band lie in the $\Gamma$ point.
Bulk silicon is the indirect-band semiconductor, with the 6 minima
of the conduction band located at between 
$\Gamma$ and $X$ points of the Brillouin zone (the distance between minima 
and $\Gamma$ point is 0.85 of the $\Gamma$--$X$ distance). 
Bulk germanium is also an indirect-band semiconductor with 4 minima of 
the conduction band in $L$ points. To consider the tunneling of 
electrons into the SiO$_2$ matrix, it is critical to 
reproduce exactly the energy positions of the edges of the conduction bands 
of the SiO$_2$ in $X$ and $L$ points.
The distances from the top of the valence band at the point $\Gamma$ to the 
points $X$ and $L$ of the conduction band for the virtual crystal are 9.11~eV 
and  8.31~eV, respectively, which is close to values in $\beta$-cristabolite.
The comparison of the band structures of the virtual crystal and $\beta$-cristobalite 
is presented in Ref.~\onlinecite{Belolipetskiy}. 
We set the top of the valence band of the bulk silicon at the energy 
distance 4.5~eV above the top of the valence band in virtual crystal.
This corresponds to the experimental data on the valence band offset 
between  bulk silicon and amorphous SiO$_2$.\cite{Sze}

\section{Results of modeling the SiGe nanocrystal in SiO$_2$}

\begin{figure}
\centering{\includegraphics[width=0.7\linewidth]{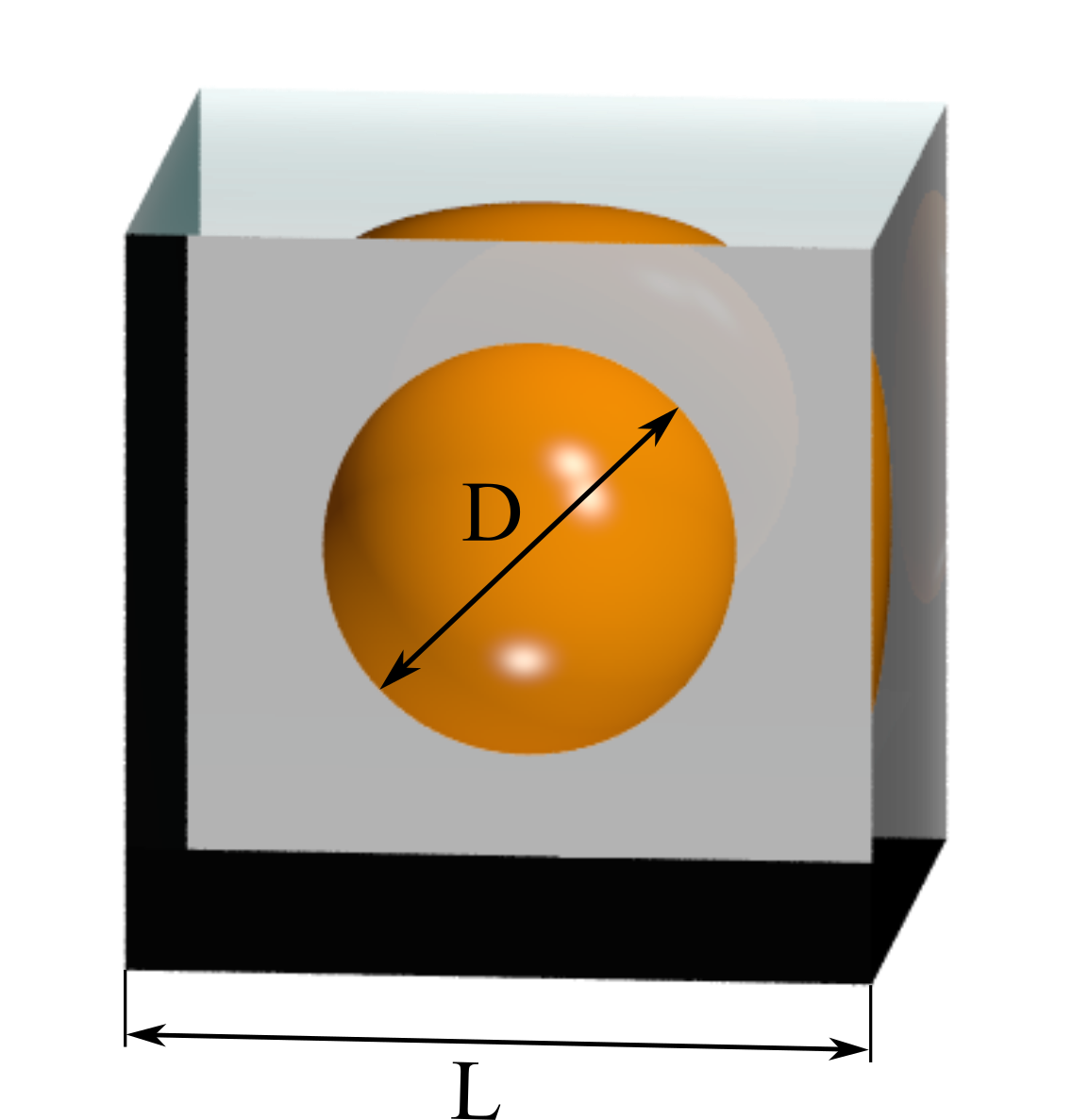}}%
\caption{
  The model of nanocrystal in SiO$_{2}$ is used in the calculation. 
  $D$ is the diameter of nanocrystal, $L$ is size of the supercell  
  with virtual SiO$_{2}$ matrix.
}\label{fig:QD_ill}
\end{figure} 

\begin{figure}
\centering{\includegraphics[width=\linewidth]{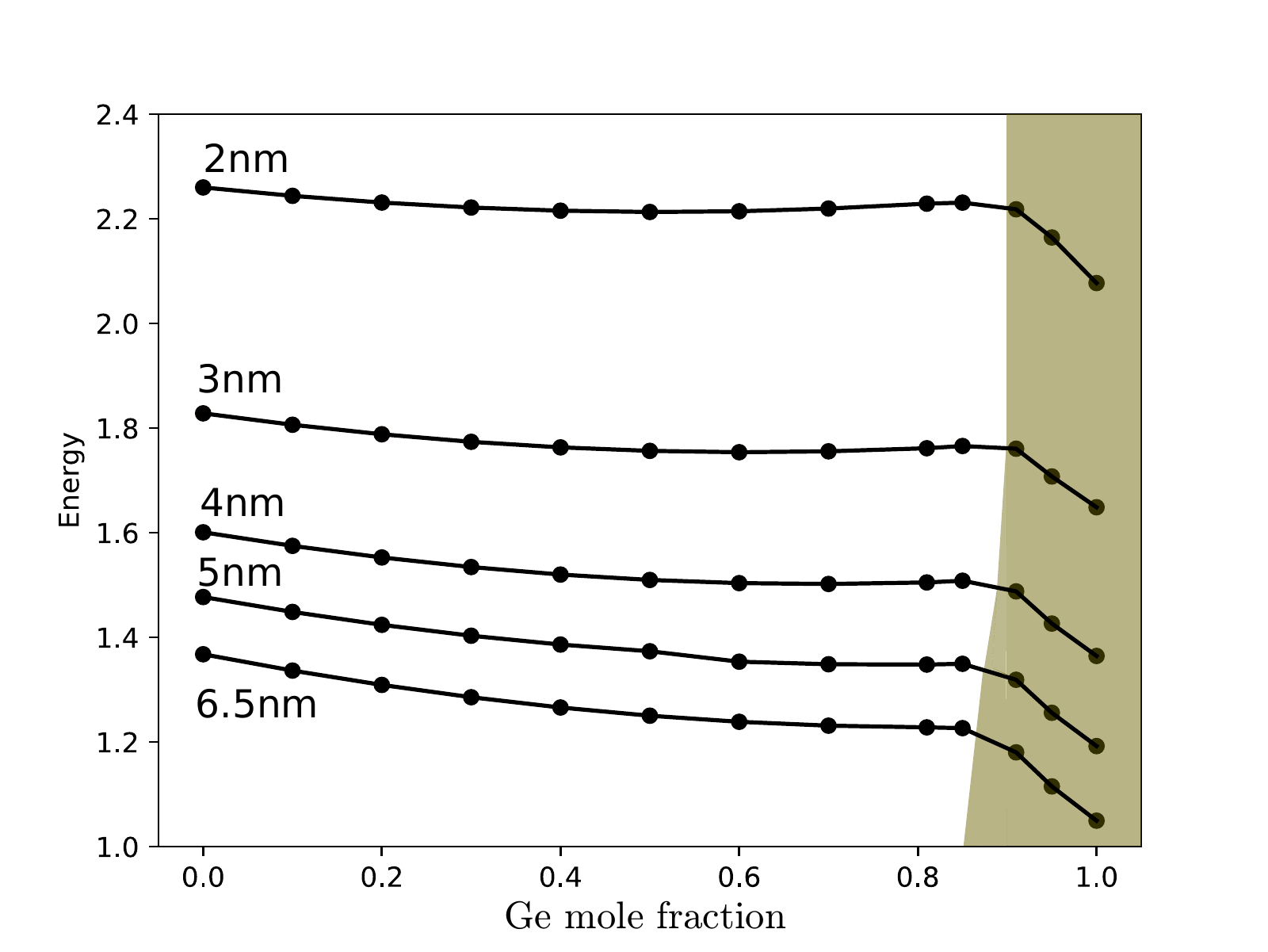}}%
\caption{
 Band gap of a SiGe NC as a function of Ge mole fraction in the SiO$_{2}$ 
 matrix for the NCs with diameter $D=2$, 3, 4, 5 and $6.5$~nm. 
 Shaded area shows NCs with the minimum of the conduction band in $L$ valley.
}\label{fig:Eg}
\end{figure}

For convenience, in the calculations of electron and hole states in 
Si$_{1-x}$Ge$_{x}$ nanocrystals embedded in the SiO$_{2}$ matrix,
we consider a cubic supercell of the virtual SiO$_2$ with a Si$_{1-x}$Ge$_{x}$ 
alloy nanocrystal in the supercell center (see Fig.~\ref{fig:QD_ill}). 
In calculations, we use the periodic boundary conditions to discard the 
effects at the surface and choose the SiO$_{2}$ cell to be large enough to 
neglect the tunneling between neighboring NCs. 
The NC with diameter $D$ is constructed by placing the SiGe ``atoms''
inside the sphere with diameter $D+0.5$~nm. Extra 0.5~nm are 
added to compensate for the interface effects, similar to 
Ref.~\onlinecite{Seino10}.
Results of calculation of the energy gap for Si$_{1-x}$Ge$_{x}$ 
nanocrystals with diameters = 2, 3, 4, 5 and 6.5~nm as a function of 
Ge content $x$ are presented in Fig.~\ref{fig:Eg}. 
The value of NC diameter corresponds to the pure silicon nanocrystal (x=0).
For the Si$_{1-x}$Ge$_{x}$ alloy nanocrystals we consider the NCs with the 
same number of atoms (the diameter is larger by the factor 
$a_{Si_{1-x}Ge_{x}}/a_{Si}$).

\section{Discussion}
First of all, let us discuss a good agreement between the detailed model 
of the SiGe alloy obtained as a averaged over the realizations of the 
random distribution of both atoms in a large supercell and the simple 
virtual crystal approximation. In addition to simplicity, the 
VCA allows for much easier analysis of the band structure behavior, as 
the bulk states do not need the extra unfolding procedure to get 
the valley index of the states. In particular, this allows one to 
unambiguously attribute the shoulder in Fig.~\ref{fig:VCA} to the 
crossover between lowest $X$ valley in SiGe alloy with low Ge content 
to the lowest $L$ valley in SiGe alloy with high Ge content.

The detailed analysis of the SiO$_2$ band structure allowed us to 
construct the tight-binding parameters of the virtual crystal which 
reproduces all important features of the $\beta$-cristabolite band 
structure. This allows one to use the atomistic tight-binding to 
compute the states in relatively large nanocrystals with the account 
on the tunneling of the states in the matrix and the valley mixing.
We stress that in this case the advantage of the tight-binding 
method is not the detailed description of the interface 
properties and/or the chemistry of the contact between SiGe 
and SiO$_2$ which is out of the scope of present paper, but the 
detailed quantitative description of the band structure of all 
materials in the full Brillouin zone and exact (within the model) 
account on the interaction between the states in different valleys. 

Calculations show that the band structure of the SiGe alloy is 
qualitatively reflected in the quantum quantization of the 
states in nanocrystals. However, there are some changes. 
First, the position of $X$--$L$ crossover point is shifted 
towards high Ge content in small NCs, see Fig.~\ref{fig:Eg}.
Second, for small NCs the band gap dependence on the Ge content 
is almost absent. This may be explained by the opposed influence 
of the badgap change and the change of the effective mass. 
As a result, for intermediate size NCs the effect of the Ge 
content on the NC band gap is strongly suppressed, and the 
band gap for 2-3~nm size NCs is constant when the Ge content 
is within the range 0.0-0.85. We remind that the NCs in 
our calculations are nominally unstrained.

\begin{figure}
\centering{\includegraphics[width=\linewidth]{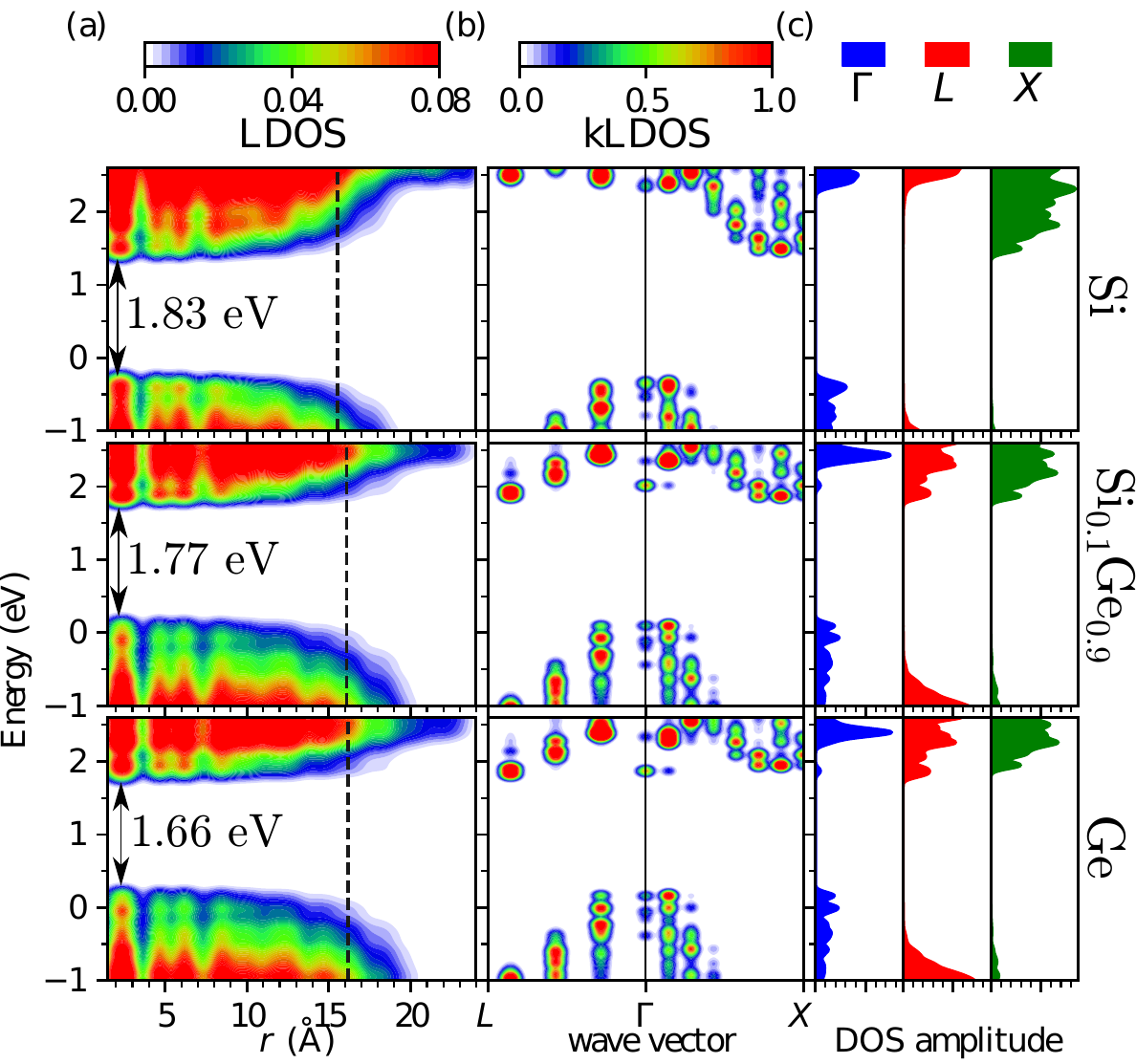}}
\caption{
 Destribution of local electron density of state in {\bf r} (a) and {\bf k} -space 
 (b) for a 3~nm size SiGe NC with the different Ge mole fraction.
}\label{fig:kLDOS}
\end{figure}

In Fig.~\ref{fig:kLDOS} we demonstrate the local density of states 
(LDOS) for electron and hole states in real and k-spaces for nanocrystal with 
diameter 3~nm (number of atoms in the NC is $1099$).

From the LDOS of the electron states 
it is easy to note that the close position of the conduction band 
in the matrix leads to the strong penetration of the hot electron 
states in the matrix. The interesting feature of the tunneling into 
matrix is that it is more ``smeared'' for Ge nanocrystals. Most 
likely, this is due to the fact that in Ge $\Gamma$ valley lies 
less than 300~meV above the conduction band bottom in $L$ valleys.
This means that electron states inside NCs with energies 
300~meV above the ground level have strong admixture of the $\Gamma$ 
valley, the same valley which forms the conduction band in the 
matrix and their tunneling is proportional to the difference 
between the energy of the state and the bottom of the conduction 
band in the matrix. As a result, the tunneling exponent is linear with 
the energy. 
However, for Si NCs the $\Gamma$ valley in the NC material lies high 
above, so the tunneling into matrix is defined by the $\Gamma$-$X$ 
mixing at the interface which is small. Only when the energy of the 
electrons reaches the conduction band of the matrix, the electron 
starts to move freely. In Fig.~\ref{fig:kLDOS} this is easy to 
see in first column: the states in purely Si NCs almost fully 
confined within the NC, and become completely delocalized 
as long as the energy of the conduction band in the matrix is 
reached. In NCs with high Ge content, the states start to 
penetrate the matrix more and more as the energy reaches the 
bottom of the matrix conduction band and even when this energy 
is reached they are still somewhat localized near the NC as long 
as they strongly feel the potential of the NC. 

In second column of Fig.~\ref{fig:kLDOS} LDOS in k-space 
shows that the states follow the band structure of the alloy. 
Distinct localization of the states near the k points of 
Brillouin zone corresponding to the position of $L$ $\Gamma$ and 
$X$ valleys for Ge NCs, and only $X$ valleys for Si NC. Note that the 
contribution to the $\Gamma$ valley for Si NCs is not from 
the Si itself, but due to states in the SiO$_2$ matrix: 
the $\Gamma$ valley in Si lies significantly above the 
maximum energy shown in this figure. 

Total valley-resolved DOS is shown in the three right columns 
in Fig.~\ref{fig:kLDOS} also shows that for Si NCs the 
states are in the $X$ valley, hot electrons with energies 
about 1.2~eV above the ground electron state acquire 
significant contribution from the $L$ valley of Si and 
$\Gamma$ valley of the SiO$_2$ matrix. 
For Si$_{0.1}$Ge$_{0.9}$ $L$ valley reaches $X$ valley, 
so the states become mixed with the $\Gamma$ valley 
for rather small energy,
but they contribute only slightly to DOS due 
to the absence of valley degeneracy and rather light 
electron mass and for purely Ge NC $L$ valley is lowest, 
with $\Gamma$ and $X$ valleys close.

\section{Conclusion}
In conclusion, we show that the VCA description in the 
framework of empirical tight-binding method for SiGe
alloys and SiO$_2$ is an effective and quantitatively 
correct approach. The calculations of electron states 
in the SiGe nanocrystals in SiO$_2$ matrix shows reach 
valley structure of the states. For Ge-rich NCs we 
demonstrate the importance of all three valleys and 
strong tunneling of excited electron states in the 
matrix. For Si-rich NCs, the states are predominantly 
$X$-valley, but the hot electron states after some 
threshold become delocalized and these delocalized 
states in the matrix and the states inside the NC 
are weakly mixed by the $\Gamma$--$X$ mixing at 
the interface.

The authors acknowledge the financial support from 
RFBR grant 18-52-54002
and the Presidium of the Russian Academy of Sciences, program no. 31.

\bibliography{TB_SiGe}

\appendix

\section{Tight-binding parameters}

\begin{table}[tbh]
\caption{Tight binding parameters of ``SiO$_2$''. Parameters SiO$_2$ are chosen 
to successfully reproduse the band stucture of $\beta$-crystobalite.}	
	\label{tbl:TB_par}
	\centering\begin{tabular}{lr}
		\hline
		 Parameters    &      ``SiO$_2$'' \\
		\hline
		$              a$ & $    5.4300$  \\
		$         E_{s}$ & $  -6.0227$  \\
		$       E_{s^*}$ & $   18.9394$  \\
		$          E_{p}$ & $    2.6548$  \\
		$         E_{d}$ & $   14.3016$  \\
		\hline
		$       ss\sigma$ & $   -2.4997$  \\
		$   s^*s^*\sigma$ & $   -3.1351$  \\
		$ ss^*\sigma$ & $   -1.7484$  \\
		$   sp\sigma$ & $    3.9755$  \\
		$ s^*p\sigma$ & $    3.0802$  \\
		$   sd\sigma$ & $   -1.3493$  \\
		$ s^*d\sigma$ & $   -4.5316$  \\
		$       pp\sigma$ & $   4.6188$  \\
		$          pp\pi$ & $   -1.0616$  \\
		$   pd\sigma$ & $   -2.8200$  \\
   		$      pd\pi$ & $    1.4004$  \\
		$       dd\sigma$ & $   -4.7861$  \\
		$          dd\pi$ & $   -0.1101$  \\
		$       dd\delta$ & $   -1.7869$  \\
		\hline
		$       \Delta/3$ & $    0.0000$  \\
	\end{tabular}
\end{table}

\end{document}